# Precise Neutron Lifetime Measurement with a Solenoidal Coil

Naoyuki Sumi[1], Hidetoshi Otono[2], Tamaki Yoshioka[2], Kenji Mishima[3] and Yasuhiro Makida[3]

[1]*Department of Physics, Graduate School of Science, Kyushu University, Fukuoka 819-0395, Japan*
[2]*Research Center for Advanced Particle Physics, Kyushu University, Fukuoka 819-0395, Japan*
[3]*KEK, High Energy Accelerator Research Organization, Tsukuba 305-0801, Japan*

E-mail: sumi@epp.phys.kyushu-u.ac.jp

The neutron lifetime, $\tau$ = 880.2 ± 1.0 sec [1], is an important parameter for particle physics and cosmology. There is, however, an 8.4 sec ($4.0\,\sigma$) deviation between the measured value of the neutron lifetime using two methods: one method counts neutrons that survive after some time, while the other counts protons resulting from neutron beta decay. A new method is being implemented at J-PARC / MLF / BL05 using a pulsed cold neutron beam. A Time Projection Chamber (TPC) records both the electrons from neutron beta decay and protons from the neutron-$^3$He capture reactions in order to estimate the neutron flux. Electron background signals require the largest correction and are source of uncertainty for this experiment. A solenoidal magnetic field can greatly reduce this background. The TPC drift region must be divided into three region in this case. A prototype detector was developed to study the multi drift layer TPC. The status of a study using a prototype detector is reported in this paper.

**KEYWORDS:** neutron lifetime, time projection chamber, solenoidal coil

## 1. Introduction

*1.1 Motivation*

The neutron lifetime measurement is important for particle physics and cosmology. Its lifetime is an input parameter for Big Bang Nucleosynthesis which predicts the abundance of light nuclei during the early universe. The light elements, such as helium and lithium, were produced by collisions of protons and neutrons. Therefore the total amount of generated light nuclei depends on the number of neutrons that existed at the time. Moreover, the neutron lifetime is an important parameter for the CKM (Cabibbo-Kobayashi-Maskawa) matrix element $V_{ud}$. A neutron lifetime measurement with improved accuracy is motivated for these reasons.

The average measurement of the neutron lifetime is 880.2 ± 1.0 sec [1]. There are two types of lifetime measurements which have been carried out so far. One is the bottle method [2], in which neutrons are stored in a special material or magnetic bottle, and neutrons which survive some time are counted. The other is the beam method [3] where neutrons are counted with a flux monitor while protons from neutron beta decay are counted using another detector. There is an 8.4 sec ($4.0\,\sigma$) deviation between the results from these two methods. A new type of measurement is therefore required to resolve the difference.

*1.2 A new type of beam method*

I discuss a new type of beam method which gives different systematic errors from previous experiments. In this new method, the so-called "electron counting beam method", the neutron lifetime is obtained by simultaneous measurement of electrons from $\beta$ decay and $^3$He neutron capture reactions recorded by a Time Projection Chamber (TPC). This method was originally developed by Kossakowski et al. [4]. The neutron lifetime $\tau$ is calculated by the following equation,

$$\tau = \frac{1}{\rho \sigma v}\left(\frac{S_{^3\text{He}}/\varepsilon_{^3\text{He}}}{S_\beta/\varepsilon_\beta}\right), \tag{1}$$

where $\rho$ is the $^3$He density, $\sigma$ is the cross section of neutron capture by $^3$He and $v$ is the neutron velocity. Since the cross section is inversely proportional to neutron velocity we use the thermal velocity $v_0 = 2200$ m/s and cross section $\sigma_0 = 5333 \pm 7$ barn for all neutron velocities. $S$ and $\varepsilon$ are the number of signals and the selection efficiency for each reaction, respectively. We aim to measure the neutron lifetime with $O(0.1\%)$ ($\sim 1$ sec) accuracy using this method.

*1.3 The largest correction*

The largest two corrections for $\tau$ are the estimation of scattered background neutrons and selection efficiency for $\beta$ decay electrons. A portion of the neutrons in beam are scattered by the TPC operation gas and captured by the detector material. Prompt gamma rays emitted from these materials produce Compton electrons. They have space, energy and time distributions similar to $\beta$ decay electrons. Therefore, we cannot reject them by signal selection nor discriminate using the method of time of flight. Lithium fluoride ($^6$LiF) suppresses the number of emitted prompt gamma rays to 0.01%. However it is estimated by Monte Carlo simulation that these backgrounds still comprise 5% of the $\beta$ decay events, which is not negligible. We can improve the purity of the $\beta$ decay signal by adopting a tighter selection process, but larger efficiency corrections are required. In any case, minimizing corrections is essential for precise measurement.

## 2. Methodology

Fig.1 shows a schematic view of this experiment. We produce a uniform magnetic field along the neutron beam axis to separate $\beta$ decay electrons from background [5]. The magnetic field enables us to separate the signal on the beam axis from the background on the wall. Also, back scattering on the TPC and the vacuum chamber surface is a leading source of inefficiency. Better signal efficiency and a lower level correction requirement can be realized by decreasing these backgrounds. We adopt a superconducting solenoidal coil which was originally prepared for the BESS experiment [6]. The drift direction of the TPC is vertically upward.

## 3. Performance

Background reduction performance was evaluated with a Monte Carlo simulation based on Geant4 [7]. Fig.2(a) shows particle tracks in the vacuum chamber as a projection to the orthogonal plane to the beam axis. Left (right) figures of Fig.2(a) correspond to $\beta$ decay (background) with and without magnetic field. The central boxes indicate the signal region. One can see that all $\beta$ decay tracks remain in the region (left bottom plot), on the other hand, only few background tracks remain (right bottom plot). Fig.2(b) shows that background is suppressed to a few % compared to the case without the field. Moreover, the magnetic field recovers the $\beta$ decay signal efficiency, because $\beta$ decay electrons do not the reach the inner wall. A magnetic field of 400 mT is enough to decrease the correction size to $O(0.1\%)$ for the neutron lifetime measurement.

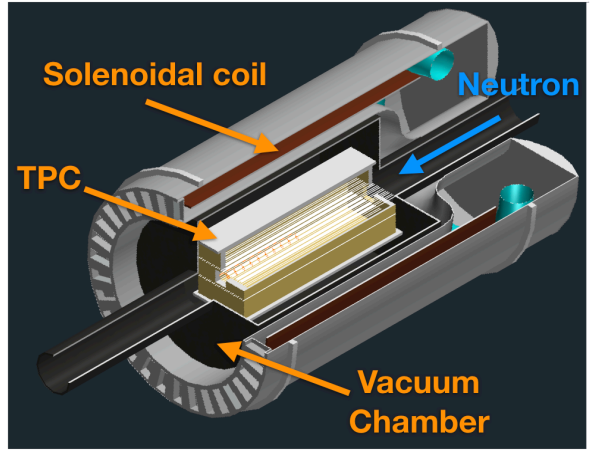

**Fig. 1.** A schematic view of the experiment. The vacuum chamber and Time Projection Chamber are located in the solenoidal coil. The neutron beam passes through this setup.

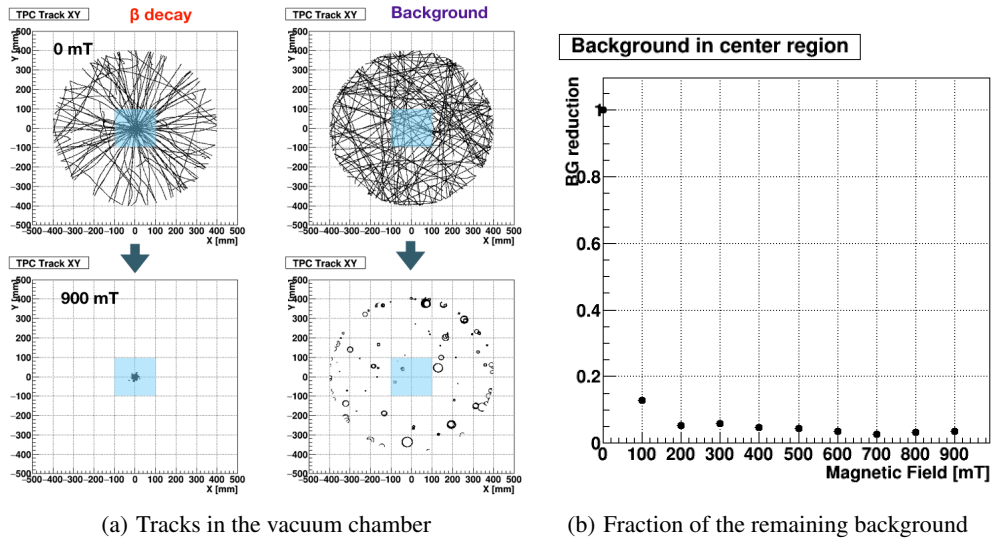

(a) Tracks in the vacuum chamber  (b) Fraction of the remaining background

**Fig. 2.** Background reduction performance with the solenoidal magnet was evaluated by Monte Carlo simulation. (a) Tracks in the vacuum chamber with and without magnetic field (900 mT). Central box indicate signal region for neutron $\beta$ decay signals. (b) Background reduction performance in the central signal region.

## 4. Prototype Detector

### 4.1 Production

A multi layered TPC is required to discriminate the central signal from other backgrounds. Since it has a unique drift field, a small prototype detector was produced in advance. A 3D CAD drawing of the detector is shown in Fig.3(a). Aluminum frames were produced at a factory in Kyushu University. Wire mounting circuits were produced by an etching method. Chip condensers and resistances were soldered on the circuits. Drift and anode wires (Be-Cu $\phi 100$ $\mu m$ and Au-W $\phi 30$ $\mu m$) were mounted to the circuits. Wire tension (100 g) was measured by sound frequency of the wires when picked with a pair of tweezers. Drift and anode voltage application cables were connected to each circuit. The constructed prototype detector is shown in Fig.3(b).

*4.2 Voltage application test*

A voltage application test of the drift and anode wires was performed in 1 atm P10 gas. We were able to apply enough voltage (1900 V) to the anode wires. Discharge between the bottom wire and the frame has been observed at 1400 V for a drift frame. We found discharged points by eye through an acrylic flange. Improved drift frames are being constructed to achieve a target voltage 3000 V.

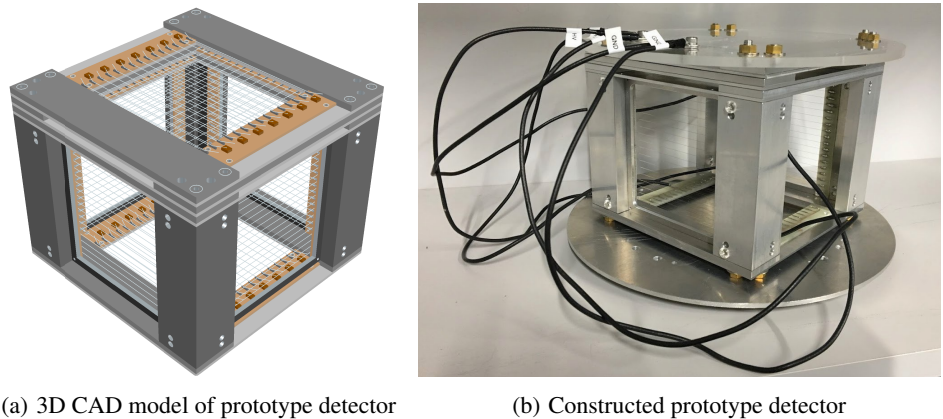

(a) 3D CAD model of prototype detector    (b) Constructed prototype detector

**Fig. 3.** A prototype detector was constructed for test of the unique drift field. Another layer of the detector is under construction for this test.

## 5. Summary and prospect

We aim to measure the neutron lifetime with $\mathcal{O}(0.1\%)$ (∼ 1 sec) accuracy using the electron counting beam method. The background in the central signal region can be suppressed using a solenoidal magnetic field. The two largest corrections can be decreased to $\mathcal{O}(0.1\%)$ of the neutron lifetime. A prototype detector was produced, and a voltage application test and a read out test are ongoing. A vacuum chamber and a full scale detector is planned to be prepared in this fiscal year.

**References**

[1] C. Patrignani et al. (Particle Data Group), Chin. Phys. C, **40**, 100001 (2016) and 2017 update.
[2] A. P. Serebrov et al., "Neutron lifetime measurements using gravitationally trapped ultracold neutrons", Phys. Rev. C **78**, 035505 (2008).
[3] J. S. Nico et al., Measurement of the neutron lifetime by counting trapped protons in a cold neutron beam, Phys. Rev. C **71**, 055502 (2005).
[4] K. Schreckenbach, G. Azuelos, P. Grivot, R. Kossakowski, and P. Liaud, Neutron de- cay measurements with a helium-filled time projection chamber, Nuclear Instruments and Methods in Physics Research A **284**, 120 (1989).
[5] Nuclear Instruments and Methods in Physics Research Section A (Volume **845**, 2017, Pages 278-280) H. Otono "LiNA - Lifetime of neutron apparatus with time projection chamber and solenoid coil", arXiv:1603.06572 [physics.ins-det]
[6] IEEE Transactions on Applied Superconductivity (Volume: 5, Issue: 2, June 1995) Y. Makida et al. "Ballooning of a thin superconducting solenoid for particle astrophysics"
[7] S. Agostinelli et al., Nuclear Instruments and Methods A **506** (2003) 250-303 "Geant4 - A Simulation Toolkit"